\documentclass[a4paper,11pt]{article}
\usepackage{pos}
\usepackage{subfig}

\title{Parametrization of the Relative Amplitude of Geomagnetic and Askaryan Radio Emission from Cosmic-Ray Air Showers using CORSIKA/CoREAS Simulations (ICRC2021)}
 \ShortTitle{Parametrization of the Relative Amplitude}

\author*[a]{Ek Narayan Paudel}
\author[a]{Alan Coleman}
\author[a,b]{Frank G. Schroeder}

\affiliation[a]{University of Delaware, Department of Physics and Astronomy, Bartol Research Institute\\
 Newark, Delaware, USA}

\affiliation[b]{Institute for Astroparticle Physics, Karlsruhe Institute of Technology, Karlsruhe, Germany} 

\emailAdd{narayan@udel.edu}
\emailAdd{alanc@udel.edu}
\emailAdd{fgs@udel.edu}

\abstract{Cosmic rays are messengers from highly energetic events in the Universe. These rare ultra-high-energy particles can be detected efficiently and in an affordable way using large arrays of radio antennas. Linearly polarized geomagnetic emission is the dominant emission mechanism produced when charged particles in air showers get deflected in the Earth’s magnetic field. The sub-dominant Askaryan emission is radially polarized and produced due to the time-varying negative-charge excess in the shower front. The relative amplitude of these two emission components depends on various air shower parameters, such as the arrival direction and the depth of the shower maximum. We studied these dependencies using CoREAS simulations of the radio emission from air showers at the South Pole using a star-shaped antenna layout. On the one hand, the parametrization of the Askaryan-to-geomagnetic ratio can be used as input for a more accurate reconstruction of the shower energy. On the other hand, if measured precisely enough, this ratio may provide a new method to reconstruct the atmospheric depth of the shower maximum.}

\FullConference{37$^{\rm{th}}$ International Cosmic Ray Conference (ICRC 2021)\\
		July 12th -- 23rd, 2021\\
		Online -- Berlin, Germany}

\begin{document}
\maketitle

\section{Introduction}\label{intro}
Energetic cosmic rays entering the Earth's atmosphere generate extensive air showers which can be detected on the ground. The charged particles, mostly e$^\pm$, in the air shower also produce coherent radio emission which can be detected by antennas at the ground level \cite{Huege:2016veh,Schroder:2016hrv}. There are two major emission mechanisms in the atmosphere: geomagnetic emission and Askaryan emission. 

\begin{figure}[h]
\centering
\includegraphics[width=9cm]{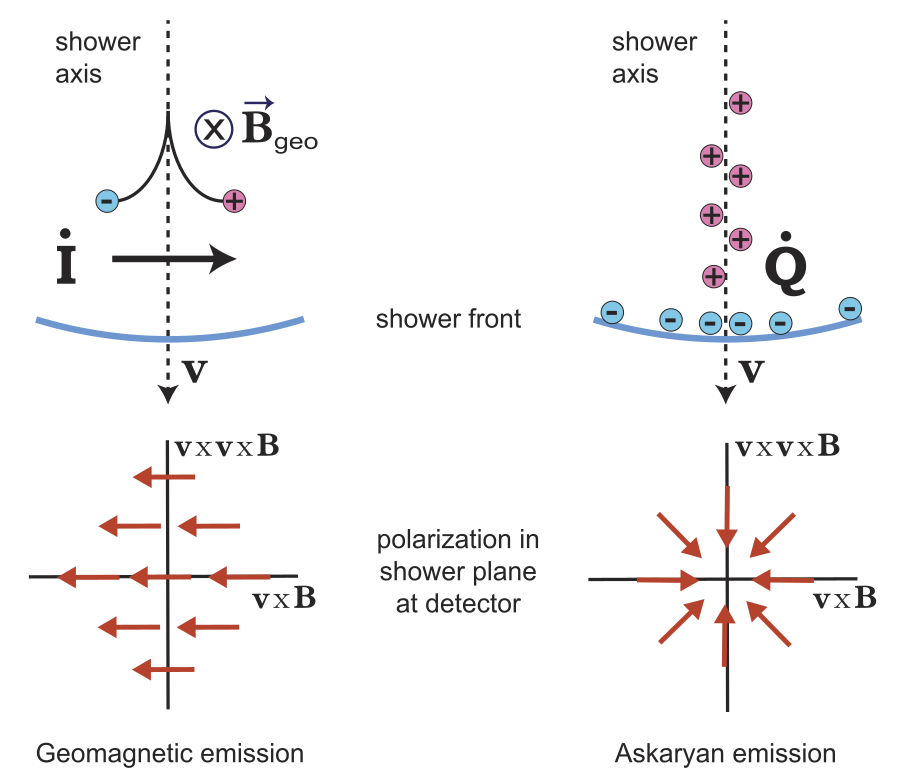}
\caption{Different radio emission mechanisms in cosmic ray air shower (from \cite{Schroder:2016hrv}). The left plot shows the separation of charge in the geomagnetic field, $\Vec{B}_{\rm geo}$, producing a transverse current, $i$. This emission is polarized along geomagnetic Lorentz force ($\Vec{v}\times\Vec{B}$ direction). The right plot shows the separation of charge along the shower axis producing Askaryan emission. It is radially polarized around the shower axis.} 
\label{fig:emissionMechanism}
\end{figure}

Geomagnetic emission is produced when the charged particles in the air shower get deflected in the Earth's magnetic field resulting in a transverse current in the direction of the geomagnetic Lorentz force. Hence, this emission is linearly polarized along $\Vec{v}\times\Vec{B}$ direction, where $\Vec{v}$ gives the direction of air shower, $\Vec{B}$ gives the direction of the geomagnetic field, as shown in the left plot in Figure \ref{fig:emissionMechanism}. Being a product of the Lorentz force, the magnitude of the emission is proportional to $\sin \alpha$ where $\alpha$ is the angle between $\Vec{v}$ and $\vec{B}$. 

Askaryan or \emph{charge excess} emission is produced due to the separation of charge along the shower axis creating the time varying negative charge excess at the shower front. This emission is radially polarized around the shower axis as shown in the right plot of Figure \ref{fig:emissionMechanism}. Unlike geomagnetic emission, Askaryan emission does not depend on the trajectory of the air shower relative to the Earth's magnetic field.

The total radio emission of an air shower is the superposition of the geomagnetic and Askaryan processes, each having different polarization directions and strengths. The constructive and destructive interference at different locations in the shower plane creates an asymmetric footprint on the ground. However, the amplitude and polarization of the radio emission measured at specific locations can be used to separate the geomagnetic and Askaryan components as already confirmed by radio experiments like LOFAR \cite{Corstanje:2015icrc}, Tunka-Rex \cite{Kostunin:2015taa} and AERA \cite{PierreAuger:2014ldh}.  

The depth of shower maximum, $X_{\rm max}$, is a mass-sensitive parameter and is from where most of the radio emission originates. The ratio of the two emission processes also has a dependence on the slant depth from the observation point to the shower maximum, $dX_{\rm max}$. In \cite{Glaser:2016qso}, a dependence of the ratio on the air density at the shower maximum has been described. Yet, it has not been studied to what extent a measurement of the radio at ground can serve as a possible method to reconstruct $X_{\rm max}$, which is done here for the location of IceCube at the South Pole. 

The surface component of the IceCube Neutrino Observatory at the South Pole, known as IceTop, is planned to be enhanced by radio antennas sensitive in the 70-350 MHz band \cite{roxmarieProceeding} and a prototype station has already succeeded in measuring air showers using radio \cite{hrvojeAlanMarieProceeding}.
We have simulated cosmic-ray air showers at the location of the South Pole and used the same frequency band of 70-350 MHz as the IceTop enhancement for our analysis which then will be applicable to such radio experiments.

In this paper, we are going to present the simulation study of the relative amplitude of two radio emission mechanisms in the air shower and its dependence on various shower parameters, such as the geomagnetic and zenith angle and $dX_{\rm max}$. We develop a model of this dependency to find the accuracy to which $X_{\rm max}$ can be determined using only the polarization of the radio emission on the ground. In section \ref{sim}, the simulations used for the analysis are described. In section \ref{methods}, the methods used for calculating the polarization, Askaryan fraction, and relative amplitude of the two radio emission mechanisms at various antenna locations are described. We then detail the development of a model which relates the magnitude of the two emissions with $dX_{\rm max}$.

\section{Simulation}\label{sim}
To perform this study we produced simulations of the radio emission from cosmic-ray air showers at the South Pole. CORSIKA was used to simulate the secondary particles and the CoREAS extension was used to simulate the radio emission~\cite{Huege:2013vt, Heck:1998vt}. We used an observation level of 2840\,m above sea level and a magnetic field of strength $|\vec{B}|$ = 54.58\,$\mu T$ inclined at an angle of $17.87^{\circ}$ from the zenith. We used Fluka2011 and SIBYLL 2.3d as high- and low-energy hadronic interaction models and the average April South Pole atmospheric profile. The simulations included the CORSIKA thinning algorithm (set to $10^{-6}$). There were 200 showers in each zenith bin of equal size in $\sin^{2}{\theta}$ = { 0.0, 0.1, ..., 0.9}. Random azimuth angles were chosen and energies were chosen according to, $d N/d E \propto E^{-1}$, from $10^{17.0}$ to $10^{17.1}$\,eV, for both iron and proton primaries.

\begin{figure}[h]
\centering
\includegraphics[width=8cm]{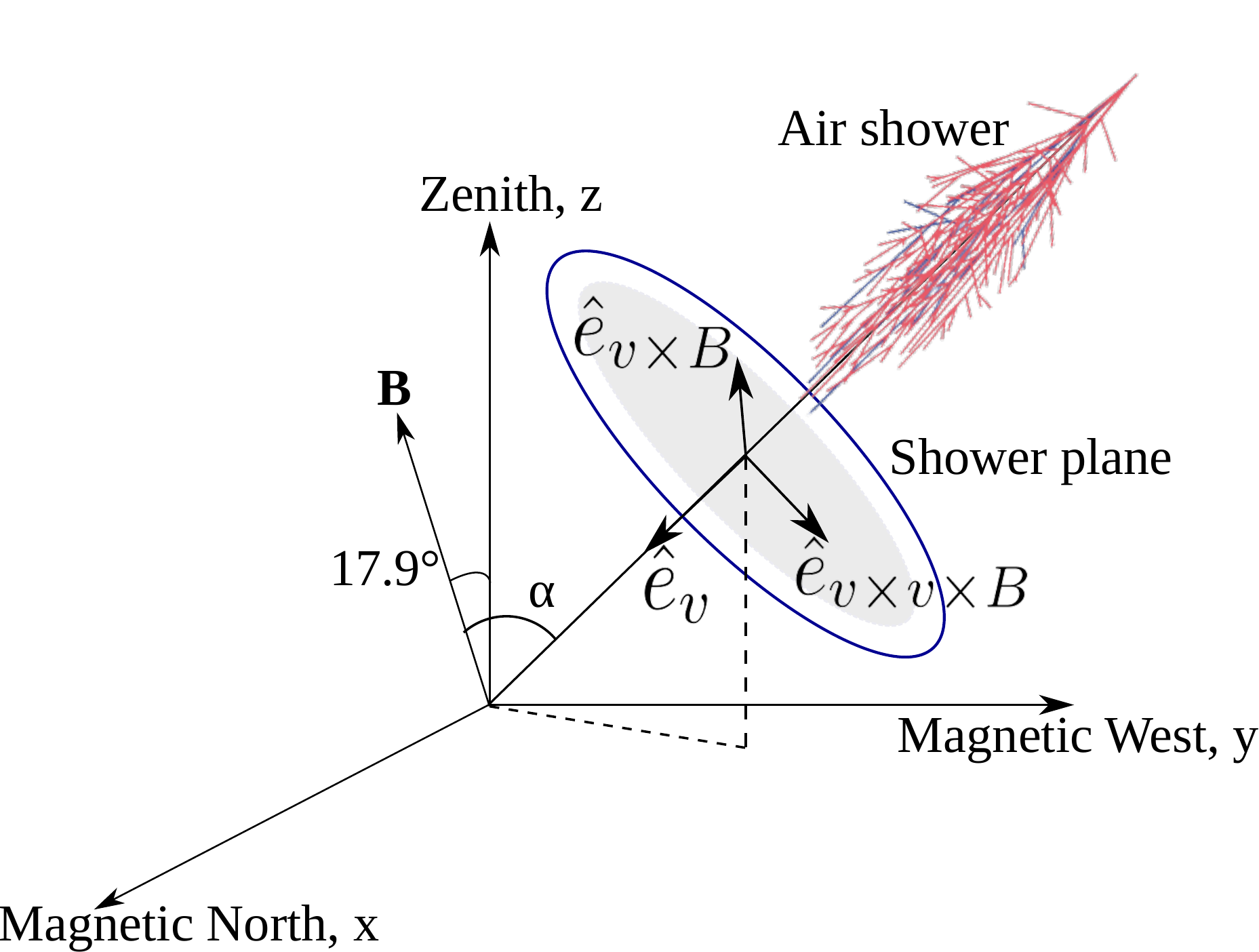}
\caption{Shower co-ordinate system used in this study with respect to the magnetic co-ordinate system.}
\label{fig:showerCoordinate}
\end{figure}

A CoREAS simulation generates the radio emission in a co-ordinate system in which the three axes are magnetic north, magnetic west and zenith. As the dominant radio emission is polarized along the Lorentz direction of the Earth's magnetic field, we instead use a shower co-ordinate system in which one axis is aligned along the Lorentz direction in the shower plane. So, the axes of shower co-ordinate systems are along $\Vec{v}\times\Vec{B}$ (x-axis), $\Vec{v}\times(\Vec{v}\times\Vec{B})$ (y-axis) and $\Vec{v}$ (z-axis), as shown in Figure \ref{fig:showerCoordinate}.
The electric fields were simulated at locations arranged in a star-shaped pattern as shown in Figure \ref{fig:antennaMapVertical}. The waveforms are then transformed into the $\vec{v}\times \vec{B}$ coordinate system. Note that in the shower coordinate system, the z axis is identical with $\Vec{v}$. Thus $E_z$ is insignificant because the hyperbolic radio wavefront is relatively flat and almost parallel to the shower plan (with a deviation of the order of $1^\circ$, only \cite{Apel:2014usa}). So, only the $x$ and $y$ components of the electric field in the shower plane are considered for relative Askaryan fraction calculation.

\section{Methods and Results}\label{methods}

\begin{figure}[h]
    \subfloat[\centering]{\includegraphics[width=7.1cm]{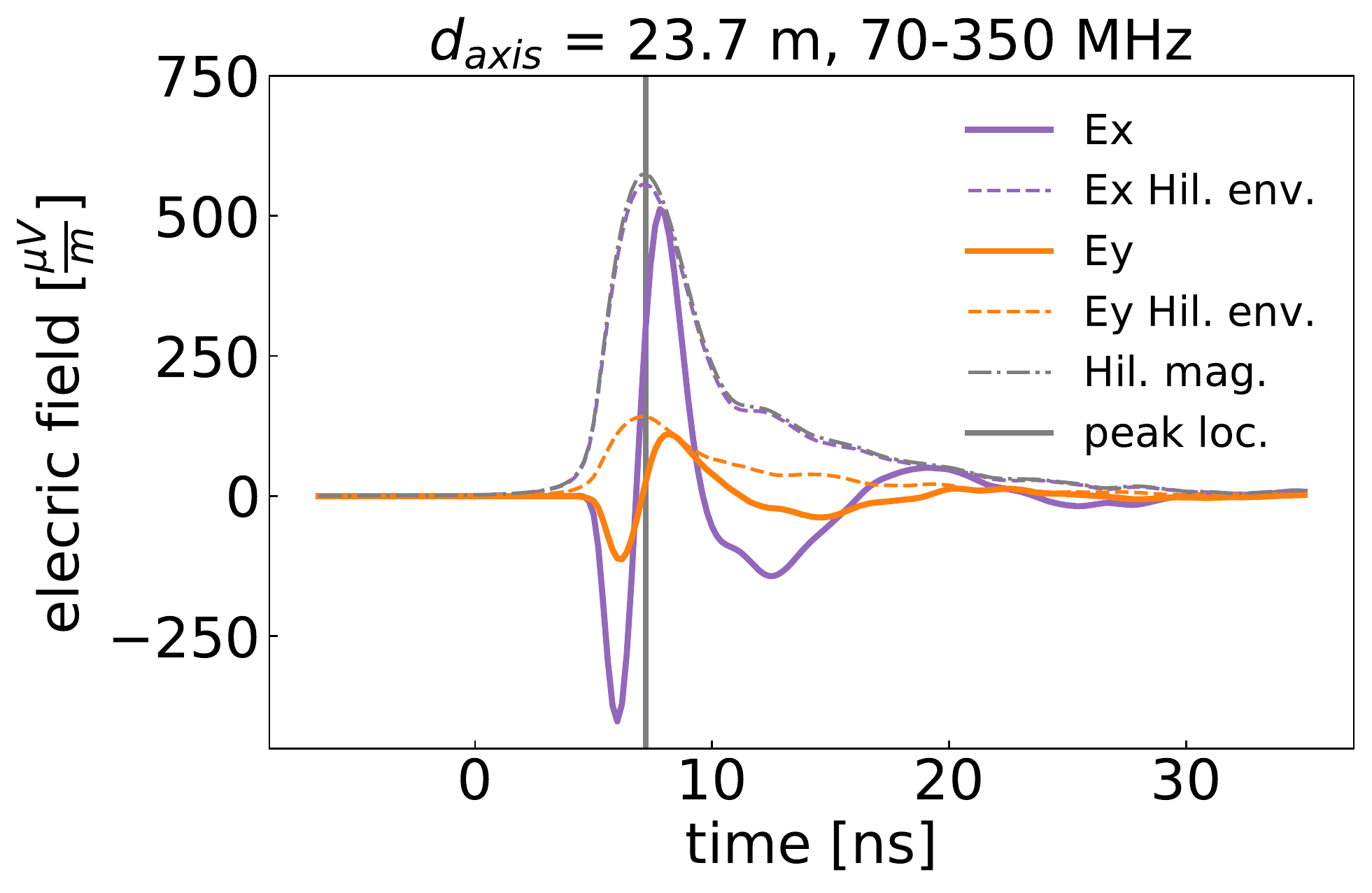}\label{fig:trace42}}
    \qquad
    \subfloat[\centering]{\includegraphics[width=7.1cm]{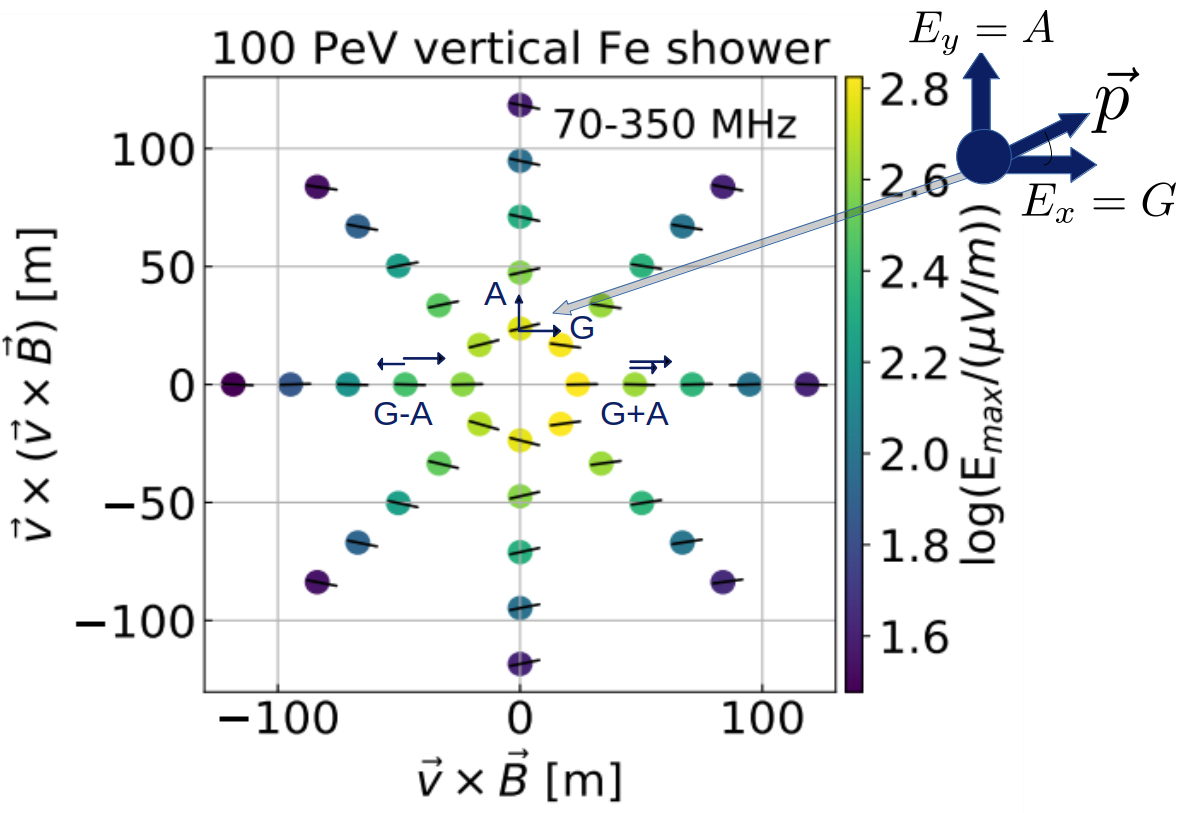}\label{fig:antennaMapVertical}}
\caption{ Figure \protect\subref{fig:trace42} shows simulated time traces and the respective Hilbert envelopes for a location along the $\Vec{v}\times(\Vec{v}\times\Vec{B})$ axis. The grey curve represents the Hilbert magnitude and the vertical line shows its peak. Figure \protect\subref{fig:antennaMapVertical} shows a map of the electric field amplitudes, given by the color, for a 100 PeV vertical iron shower. 
For locations along y-axis, the geomagnetic and Askaryan contributions are orthogonal to each other with the x-component representing the geomagnetic contribution and the y-component representing the Askaryan contribution.}
\label{fig:trace}
\end{figure}

The waveforms are filtered to the band of 70\,MHz to 350\,MHz. Figure \ref{fig:trace42} shows an example of a filtered electric field at 23.7\,m from the shower axis. We used the Hilbert envelope of the total signal to identify its peak and used the individual Hilbert components at the peak to calculate the polarization plane. 
This implies the approximation that the total radio emission is approximately linearly polarized, ignoring the phase shift between the two emission mechanisms and the resulting small circular polarization component \cite{Scholten:2016gmj}.
As shown in Figure \ref{fig:antennaMapVertical}, the two emission mechanisms are orthogonal to each other along the $\Vec{v}\times(\Vec{v}\times\Vec{B})$ axis with the geomagnetic emission in the x-direction, $G = |E_x|$, and the Askaryan in the y-direction, $A = |E_y|$. We calculate two fractions corresponding to the relative Askaryan fraction, \ref{eq:AFvB}, and the simple ratio of the geomagnetic and Askaryan amplitudes, equation \ref{eq:GAvB}.
\begin{equation}
\label{eq:AFvB}
 \frac{A}{A+G} =\frac{ \left\lvert E_{y} \right\rvert}{ \left\lvert E_{y}  \right\rvert+ \left\lvert E_{x} \right\rvert} 
\end{equation}
\begin{equation}
\label{eq:GAvB}
 \frac{G}{A} =\frac{ \left\lvert E_{x} \right\rvert}{ \left\lvert E_{y}  \right\rvert} 
\end{equation}
To estimate the uncertainty in equations \ref{eq:AFvB} and \ref{eq:GAvB} for an individual waveform, we performed a bootstrap estimation. We used the RMS (root mean square) of the last 40\% of the length of the signal trace along each axis which gives us an estimate of the noise in the signal (numerical noise and incoherent radiation; no background was added). We chose a vector with a magnitude equal to the RMS value and with a random direction and added it to the peak electric field and recalculated the fractions. This process was repeated 1000 times and the 68\% interval of the resulting distribution was used as an uncertainty estimate of the relative Askaryan fraction. The same method was also used to estimate the uncertainties of the relative amplitude, $\frac{G}{A}$.

\begin{figure}[t]
\centering
\includegraphics[width=8cm]{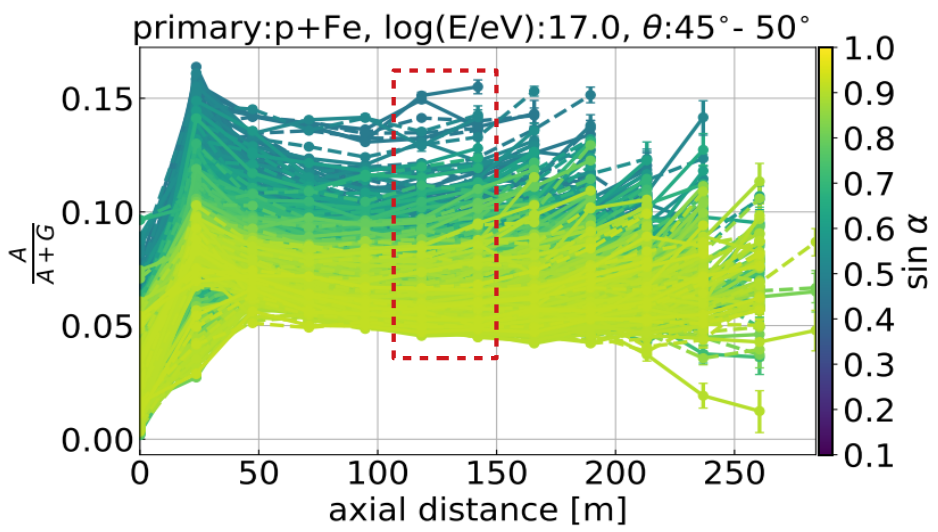}
\caption{Relative Askaryan fraction plotted against the distance to the shower axis for showers with zenith angles between 45$^{\circ}$ and 50$^{\circ}$. Solid (dotted) lines are for locations along the positive (negative) y-axis. The red dotted box represents the region in the plots where the value of the relative Askaryan fraction is at a plateau. These showers are for proton and iron primaries with energies between $10^{17.0}$ and $10^{17.1}$ eV. The colorbar represents $\sin \alpha$ with $\alpha$ being the geomagnetic angle.}
\label{fig:caseAG}
\end{figure}

In Figure \ref{fig:caseAG}, the relative Askaryan fraction (equation \ref{eq:AFvB}) is plotted against the distance from the shower axis. We used a signal-to-noise ratio (SNR) cut of SNR$ >10^{4}$ to exclude the locations where waveforms with weak, non-coherent signals are present.

\begin{equation}
\label{eq:SNR}
 \frac{S}{N} =\left(\frac{{\rm Signal}_{\rm peak}}{{\rm Noise}_{\rm RMS}}\right)^2 
\end{equation}
In equation \ref{eq:SNR}, ${\rm Signal}_{\rm peak}$ is the peak of the Hilbert envelope of the signal trace and ${\rm Noise}_{\rm RMS}$ is the root mean square of the 40\% tail of that trace. 

The relative Askaryan fraction increases with axial distance, eventually reaching a plateau at a distance from the shower axis that depends on the radius of the Cherenkov ring of the shower. The weighted average value of $\frac{A}{A+G}$ was calculated for measurements in a radial windows. The windows used in this study are given in table \ref{table:plateauWindow}.

\begin{table}[t]
\centering
\begin{tabular}{c|c}
\hline
 zenith angle $\theta ^{\circ}$ & window edges\\ 
 \hline
 <25$^{\circ}$ & 40 m - 80 m  \\ 

 25$^{\circ}$- 35$^{\circ}$ &  60 m  - 100 m \\

 35$^{\circ}$- 40$^{\circ}$ &   90 m  - 130 m \\

 40$^{\circ}$- 55$^{\circ}$ &   110 m  - 150 m\\

 55$^{\circ}$- 60$^{\circ}$ &   110 m  - 180 m\\

 60$^{\circ}$- 65$^{\circ}$ &   140 m  - 220 m\\

 65$^{\circ}$- 70$^{\circ}$ &   210 m  - 330 m\\

 70$^{\circ}$- 75$^{\circ}$ &   260 m  - 420 m\\

 75$^{\circ}$- 80$^{\circ}$ &   410 m  - 690 m\\

 >80$^{\circ}$ &   600 m  - 1060 m\\
 \hline
\end{tabular}
\caption{Table showing distances from shower axis in shower plane used to locate plateau window for relative Askaryan fraction for the showers of different zenith angles. These intervals are based on the location of the Cherenkov ring of the showers.}
\label{table:plateauWindow}
\end{table}
 
\begin{figure}[t]
    \subfloat[\centering]{\includegraphics[width=7.5cm]{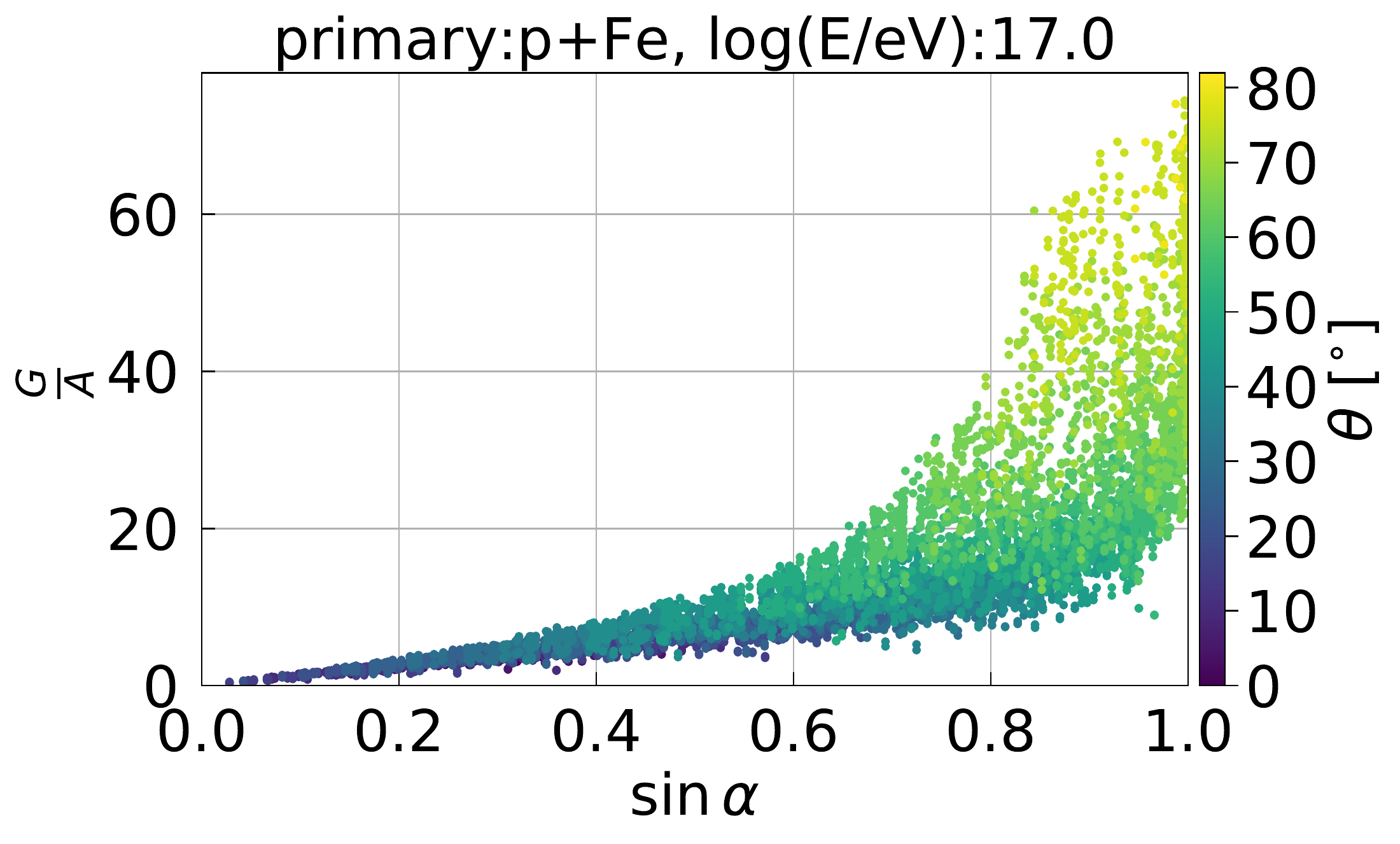}\label{fig:PlateauGASinAlpha}}\hspace*{-2em}
    \qquad
    \subfloat[\centering]{\includegraphics[width=7.5cm]{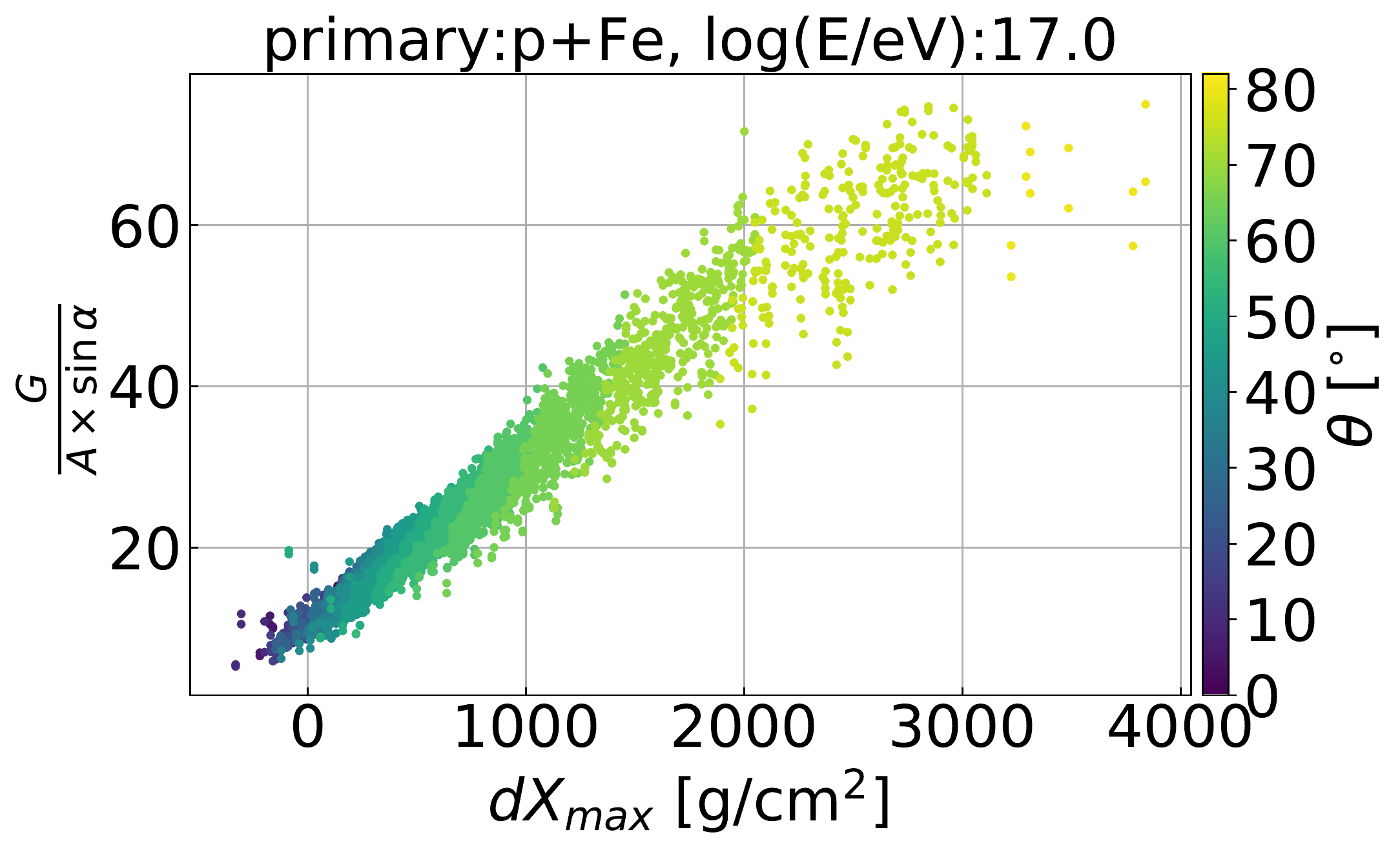}\label{fig:PlateauGAdXmax}}
\caption{\protect\subref{fig:PlateauGASinAlpha} Plateau value of $\frac{G}{A}$ plotted against $\sin{\alpha}$ for various 100 PeV proton and iron showers. \protect\subref{fig:PlateauGAdXmax} Plateau value of  $\frac{G}{A \times \sin{\alpha} }$ plotted against dX$_{\rm max}$ for various 100 PeV proton and iron showers. The colorbar represents the zenith angle of the shower.}
\label{fig:sinAlphadXmax}
\end{figure}

We mostly used the relative Askaryan fraction because it is, by definition, confined between 0 and 1 and easy to use. But to study the effect of $\sin \alpha$, which affects only the geomagnetic emission, we switched to the simple ratio of the two processes (equation \ref{eq:GAvB}). In Figure \ref{fig:PlateauGASinAlpha}, $\frac{G}{A}$ is plotted against $\sin{\alpha}$. The color-bar represents the zenith angle of the shower. The relative amplitude has a clear dependence on $\sin{\alpha}$, as expected, because of the geomagnetic emission, which theoretically goes to zero with $\sin \alpha = 0$. However, there is also a very clear dependence on the zenith angle in addition to implicit zenith dependence contained in the geomagnetic angle $\alpha$.

This additional zenith dependency is due to the distance to the shower maximum $dX_{\rm max}$.
To show this, we rescaled the Askaryan fraction to directly include the geomagnetic angle, $\frac{G}{A \times \sin \alpha}$. In Figure \ref{fig:PlateauGAdXmax}, the rescaled amplitude is shown against $dX_{\rm max}$ in units of g/cm$^2$.
A clear, nearly linear, relationship is observed. We found similar relationship of the rescaled amplitude with $dX_{\rm max}$ measured in m.

One can then use this relationship between $dX_{\rm max}$ and the more easily identified shower observables, which are the polarization direction of the electric field at various locations in the shower plane and the geomagnetic angle, to extract the mass sensitive parameter, $X_{\rm max}$.

\begin{figure}[h]
\centering
\includegraphics[width=10cm]{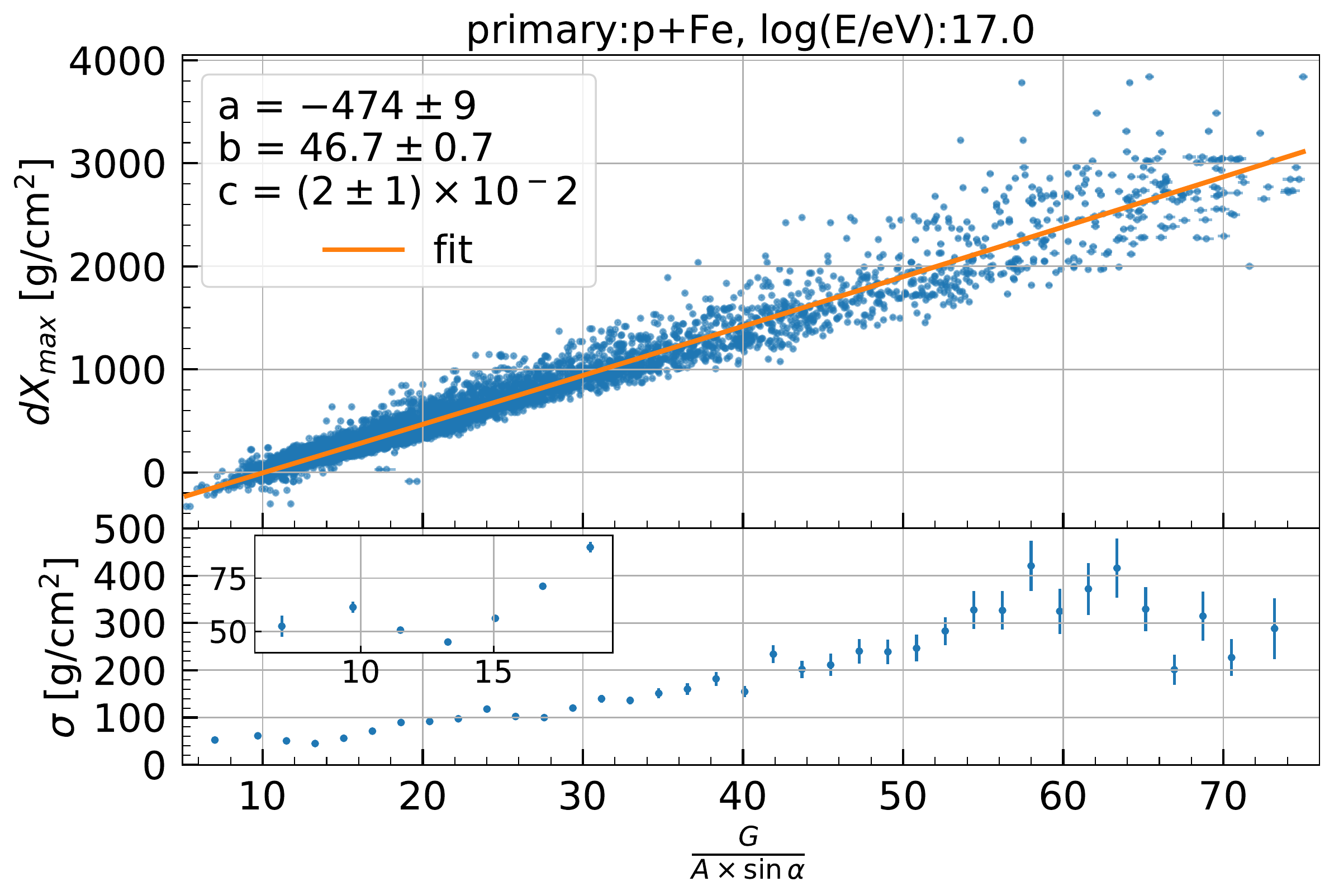}
\caption{dX$_{\rm max}$ plotted against the plateau value of $\frac{G}{A \times \sin{\alpha}}$ for proton and iron showers with energy $10^{17} - 10^{17.1}$ eV. The orange line shows a polynomial fit to the blue points representing $dX_{\rm max}$. The lower subplot shows standard deviation of the $dX_{\rm max}$ from fitted values.}
\label{fig:PlateauGADivSinAdXmaxWSD0_20}
\end{figure}

In Figure \ref{fig:PlateauGADivSinAdXmaxWSD0_20}, dX$_{\rm max}$ was plotted against the values of $\frac{G}{A \times \sin{\alpha}}$ for amplitudes up to 75 using proton and iron air shower simulations. A second degree polynomial $-474 + 46.7 x +2 x^{2}$ was fitted to the points. In the bottom portion of the figure, we show the spread of the data about the fit line. It shows that the spread between the reconstructed and true $dX_{\rm max}$ is best for showers with small values of the rescaled Askaryan fraction, $\gtrsim$50\,g/cm$^{2}$. These would correspond to more vertical and/or deep showers as indicated by the small $dX_{\rm max}$ values. For showers which reach the maximum further from the observation level, the resolution is $\sim$100\,g/cm$^{2}$, and even worse for very inclined showers. Thus, this method, alone, would not be competitive with other modern techniques to reconstruct $X_{\rm max}$, which have a precision of about 20 to 30\,g/cm$^{2}$~\cite{Buitink:2014eqa,Bezyazeekov:2018yjw,Pont:2021icrc}.

\section{Conclusions}\label{conclusions}
We studied the relative Askaryan fraction and the relative amplitude of radio emission processes from cosmic-ray air showers using CoREAS simulations for $10^{17.0} - 10^{17.1}$ eV proton and iron primaries using a star shaped layout. The relative Askaryan fraction has a dependence on both $\sin \alpha$ and $dX_{\rm max}$. This dependence was used to estimate $dX_{\rm max}$ using measured values of the relative amplitude of the Askaryan and geomagnetic emissions on the surface. This study was performed in the same band that will be used by the surface enhancement of IceTop at the South Pole. We found that the accuracy in $dX_{\rm max}$ obtained from this parametrization is not sufficient to be competitive with other methods of reconstructing the depth of shower maximum.


\bibliographystyle{ICRC}
\bibliography{main}

\providecommand{\href}[2]{#2}\begingroup\raggedright\begin{thebibliography}{10}

\bibitem{Huege:2016veh}
T.~Huege \href{http://dx.doi.org/10.1016/j.physrep.2016.02.001}{{\em Phys.
  Rept.} {\bfseries 620} (2016) 1--52}.

\bibitem{Schroder:2016hrv}
F.~G. Schr\"oder \href{http://dx.doi.org/10.1016/j.ppnp.2016.12.002}{{\em Prog.
  Part. Nucl. Phys.} {\bfseries 93} (2017) 1--68}.

\bibitem{Corstanje:2015icrc}
{\bfseries LOFAR} Collaboration, A.~Corstanje {\em et~al.} {\em PoS} {\bfseries
  ICRC2015} (2015) 396.

\bibitem{Kostunin:2015taa}
D.~Kostunin, P.~A. Bezyazeekov, R.~Hiller, F.~G. Schr\"oder, V.~Lenok, and
  E.~Levinson \href{http://dx.doi.org/10.1016/j.astropartphys.2015.10.004}{{\em
  Astropart. Phys.} {\bfseries 74} (2016) 79--86}.

\bibitem{PierreAuger:2014ldh}
{\bfseries Pierre Auger} Collaboration, A.~Aab {\em et~al.}
  \href{http://dx.doi.org/10.1103/PhysRevD.89.052002}{{\em Phys. Rev. D}
  {\bfseries 89} no.~5, (2014) 052002}.

\bibitem{Glaser:2016qso}
C.~Glaser, M.~Erdmann, J.~R. H\"orandel, T.~Huege, and J.~Schulz
  \href{http://dx.doi.org/10.1088/1475-7516/2016/09/024}{{\em JCAP} {\bfseries
  09} (2016) 024}.

\bibitem{roxmarieProceeding}
{\bfseries IceCube} Collaboration, M.~Oehler and R.~Turcotte {\em PoS}
  {\bfseries ICRC2021} (2021) 225.

\bibitem{hrvojeAlanMarieProceeding}
{\bfseries IceCube} Collaboration, H.~Dujmovic, A.~Coleman, and M.~Oehler {\em
  PoS} {\bfseries ICRC2021} (2021) 314.

\bibitem{Huege:2013vt}
T.~Huege, M.~Ludwig, and C.~W. James
  \href{http://dx.doi.org/10.1063/1.4807534}{{\em AIP Conf. Proc.} {\bfseries
  1535} no.~1, (2013) 128}.

\bibitem{Heck:1998vt}
D.~Heck {\em et~al.}, {\em Report FZKA 6019, Forschungszentrum Karlsruhe},
  1998.

\bibitem{Apel:2014usa}
{\bfseries LOPES} Collaboration, W.~D. Apel {\em et~al.}
  \href{http://dx.doi.org/10.1088/1475-7516/2014/09/025}{{\em JCAP} {\bfseries
  09} (2014) 025}.

\bibitem{Scholten:2016gmj}
{\bfseries LOFAR} Collaboration, O.~Scholten {\em et~al.}
  \href{http://dx.doi.org/10.1103/PhysRevD.94.103010}{{\em Phys. Rev. D}
  {\bfseries 94} no.~10, (2016) 103010}.

\bibitem{Buitink:2014eqa}
{\bfseries LOFAR} Collaboration, S.~Buitink {\em et~al.}
  \href{http://dx.doi.org/10.1103/PhysRevD.90.082003}{{\em Phys. Rev. D}
  {\bfseries 90} no.~8, (2014) 082003}.

\bibitem{Bezyazeekov:2018yjw}
{\bfseries Tunka-Rex} Collaboration, P.~A. Bezyazeekov {\em et~al.}
  \href{http://dx.doi.org/10.1103/PhysRevD.97.122004}{{\em Phys. Rev. D}
  {\bfseries 97} no.~12, (2018) 122004}.

\bibitem{Pont:2021icrc}
{\bfseries Pierre Auger} Collaboration, B.~Pont {\em PoS} {\bfseries ICRC2021}
  (2021) 387.

\end{thebibliography}\endgroup
\end{document}